Jawaher Almutlaq,* Kyle P. Kelley,* Hyeongrak Choi, Linsen Li, Benjamin Lawrie, Ondrej Dyck, Dirk Englund,* and Stephen Jesse†

# Closed-Loop Electron-Beam-Induced Spectroscopy and Nanofabrication Around Individual Quantum Emitters

**Abstract** Color centers in diamond play a central role in the development of quantum photonic technologies, and their importance is only expected to grow in the near future. For many quantum applications, high collection efficiency from individual emitters is required, but the refractive index mismatch between diamond and air limits the optimal collection efficiency with conventional diamond device geometries. While different out-coupling methods with near-unity efficiency exist, many have yet to be realized due to current limitations in nanofabrication methods, especially for mechanically hard materials like diamond. Here, we leverage electron-beam-induced etching to modify Sn-implanted diamond quantum microchiplets containing integrated waveguides with width and thickness of 280 nm and 200 nm, respectively. This approach allows for simultaneous high-resolution imaging and modification of the host matrix with an open geometry and direct writing. When coupled with the cathodoluminescence signal generated from the electron-emitter interactions, we can monitor the enhancement of the quantum emitters in real-time with nanoscale spatial resolution. The operando measurement and manipulation of single photon emitters demonstrated here provides a new foundation for the control of emitter-cavity interactions in integrated quantum photonics.

**Keywords:** diamond chiplets, color centres in diamond, SnV-, electron-beam-induced etching (EBIE), cathodoluminescence.

## 1 Introduction

Diamond color centers, notably the nitrogen-vacancy (NV) and group-IV color centers like the tin-vacancy (SnV-), have emerged as crucial components in quantum information technology [1]–[5]. They serve as photonic interfaces to electron and nuclear spin ground states, driving research in applications such as quantum sensing, quantum computing, and quantum networking [6]–[8]. Nevertheless, a significant challenge in this field lies in fabricating photonic nanostructures, such as waveguides and photonic crystal nanocavities, to efficiently manipulate and couple photons to the spin of color centers.[9]

The obstacle is particularly pronounced in diamond due to its unique properties, including high hardness and chemical stability. Conventional methods like electron-beam lithography (EBL) and reactive ion etching (RIE) have demonstrated remarkable progress over the past few years, enabling the development of diamond quantum microchiplets (QMCs) [10] and nanobeam waveguides[11]. However, RIE requires the development of a hard mask and the resolution is limited to 100 nm. Focused ion beam (FIB) milling, on the other hand, induces a damage layer in diamond that is 50 nm in depth for gallium sources and up to 500 nm in depth for lighter species such as oxygen; this FIB-induced damage is known to quench color center emission in diamond[12]. Femtosecond laser writing enabled the fabrication of diamond-based three-dimensional (3D) waveguides[13], [14], and micro-pillar structures[15], but the structure size is limited to the microscale and can be highly dependent on spherical aberrations[16]. Recently, electron-beam-induced etching (EBIE) has emerged as an effective etching method for patterning diamond substrates, evidenced by the low damage and absence of graphitization [17]. In this method, the ions in FIB are replaced with a focused electron beam in a water vapor environment; the electron beam doesn't etch the material directly, but it initiates a surface reaction for material removal in a relatively pure chemical environment [18], [19]. This technique has enabled direct-write 3D sculpting with a high resolution on the order of 10 nm[20], [21]. However, challenges in spatially and spectrally aligning nanoscale photonic cavities with sparsely distributed single photon emitters (SPEs) in diamond remain a substantial obstacle to the development of high quality-factor diamond quantum photonic platforms.

Here, we manipulate the optical properties of SPEs in diamond QMCs using in-situ mask-less EBIE and hyperspectral cathodoluminescence (CL) microscopy performed simultaneously (Figure 1a). While conventional optical probes of SPEs are limited by the optical diffraction limit, CL microscopy relies on a sub-nm converged electron-beam probe and offers spatial resolution limited only by secondary electron scattering and free-carrier migration. We show here that we can deterministically etch photonic cavities using EBIE around individual SPEs identified by CL. Collecting the photons from in-situ CL while etching a SnV- SPE embedded in a suspended diamond QMC waveguide, we observed up to ~2.5x emission enhancement followed by emission quenching as the SPE was etched away. These results show that the EBIE induces minimal damage, if any, and the emitter remained active until it was removed from the 200-nm thick waveguide. Further, we showed the feasibility of fabricating an array of cavities approximately 20 nm in diameter with minimal variations along the waveguide. This highly integrated approach can serve as a rapid testbed for various complex designs with in-situ testing and imaging, establishing a direct correlation between design attributes and performance metrics. While this method is not meant to address the issue of scalability or entirely replace RIE, it can nicely complement it, especially for sub-100 nm features and in-situ emitter enhancement tests. This method facilitates rapid iterations and refinements in the development process, significantly reducing the time from concept to application.

*Corresponding author: Dirk Englund, Massachusetts Institute of Technology, Cambridge, MA, United States; englund@mit.edu
†Corresponding author: Stephen Jesse, Center for Nanophase Materials Sciences, Oak Ridge National Laboratory, Oak Ridge, TN, United States; sjesse@ornl.gov
Jawaher Almutlaq, Hyeongrak Choi, Linsen Li, Massachusetts Institute of Technology, Cambridge, MA, United States
Ondrej Dyck, Kyle Kelley, Center for Nanophase Materials Sciences, Oak Ridge National Laboratory, Oak Ridge, TN, United States
Benjamin Lawrie, Materials Science and Technology Division, Oak Ridge National Laboratory, Oak Ridge, TN, United States

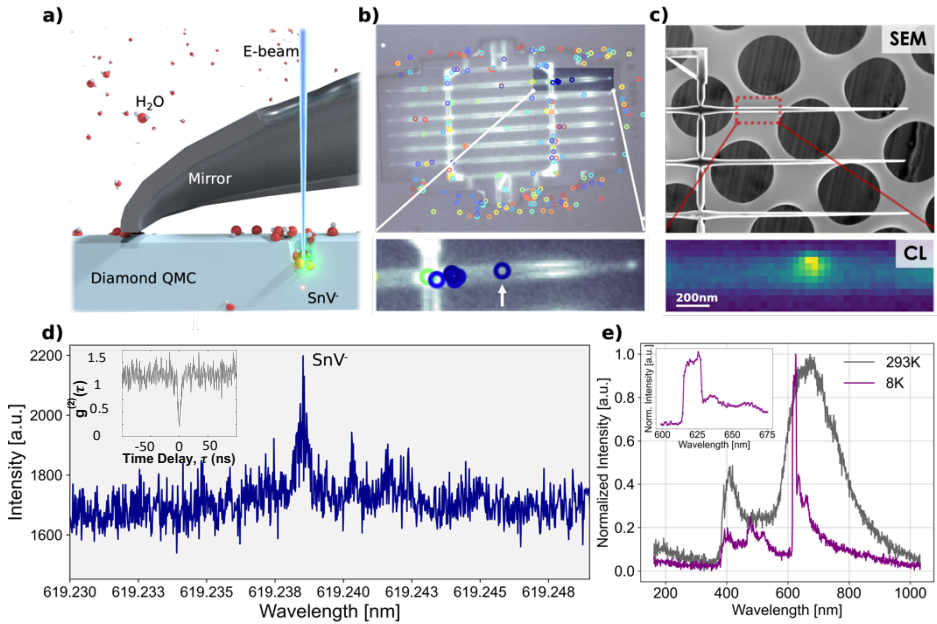

**Figure 1:** Experimental design for diamond QMC etching and analysis. (a) SEM-CL setup: converged electron beam incident on a diamond QMC through a pinhole in a parabolic mirror in an environmental SEM operating in 100 Pa water vapor environment at a temperature of 300K. (b) Diamond QMC image overlaid with the location of SPEs characterized by the widefield PLE at 4 K. Each circle denotes an SPE. (c) SEM image of a diamond QMC indicating area of interest (red box) with CL map displaying emission at 620 nm from the emitter. (d) PLE spectrum centered around 620 nm from SnV- represented in panels (b,c ; inset shows a representative autocorrelation measurement of single SnV-. (e) CL spectrum for an individual SPE at room temperature (grey) and 8 K (purple).

## 2 Principle and Experimental Methods

### 2.1 Gas-assisted EBIE and CL in diamond

The EBIE and CL microscopy was performed in an FEI Quattro environmental scanning electron microscope (SEM) using a Delmic Sparc CL module to collect luminescence generated by the electron beam into an Andor Kymera 193i spectrograph. A flip mirror was used to redirect the collected CL away from the spectrograph and into a 50/50 fiber splitter for photon correlation function measurements performed with a pair of superconducting nanowire single photon detectors and a Swabian TimeTagger Ultra used to time tag photon detection events.

The emitter was first located and characterized in no water vapor (i.e. pressures below 1E-6 Torr). Then, the EBIE was performed in a 0.75 Torr water vapor background. CL was acquired concurrently during the EBIE process, and reference CL measurements were taken at pressures of less than 1E-6 Torr, where no etching was observed even with extended electron-beam dwell times. A Gatan cryostage was used to acquire CL spectra at 8K, though, because the water vapor background is incompatible with cryogenic operation, no EBIE was observed at cryogenic temperatures. Notably, cooling the sample to 8K immediately after switching to high-vacuum operation results in ice formation on the diamond QMC at 8K, but this water reservoir is quickly removed by electron-beam exposure, and it provides an insufficient environment to facilitate cryogenic EBIE.

In this room temperature (293 K) experiment, the $H_2O$ vapor injected into the chamber adsorbs onto the diamond surface, forming bonds with the diamond surface mediated by the oxygen atoms. The incident and emitted electrons then interact with these adsorbed molecules, dissociating them and generating reactive radicals (i.e., $H^+$ and $OH^-$) that facilitate local chemical dry etching. The reaction products, such as $CO_2$ and CO, formed in this process are volatile and spontaneously desorb from the surface, leaving behind etched areas[17], [20]. Several factors, including the dose of the electron beam, the kinetics of the reaction and the nature and pressure of the injected gas, determine the spatial resolution and etch depth in EBIE. The sub-nm electron beam spot size allows for the local etching of specific structures, enabling the direct writing of a wide range of photonic cavities and structures within a single system that also serves as a photonic interface for quantum technologies. EBIE has been used with various combinations of gases and substrate materials[18], [22], including recent efforts focused on nanoscale patterning of 2D quantum nanophotonic systems[23], [24]. Although this process has traditionally been used for material etching, recent advancements have expanded its applications to material analysis, surface functionalization, and nanostructured material growth.[17], [25]–[27]

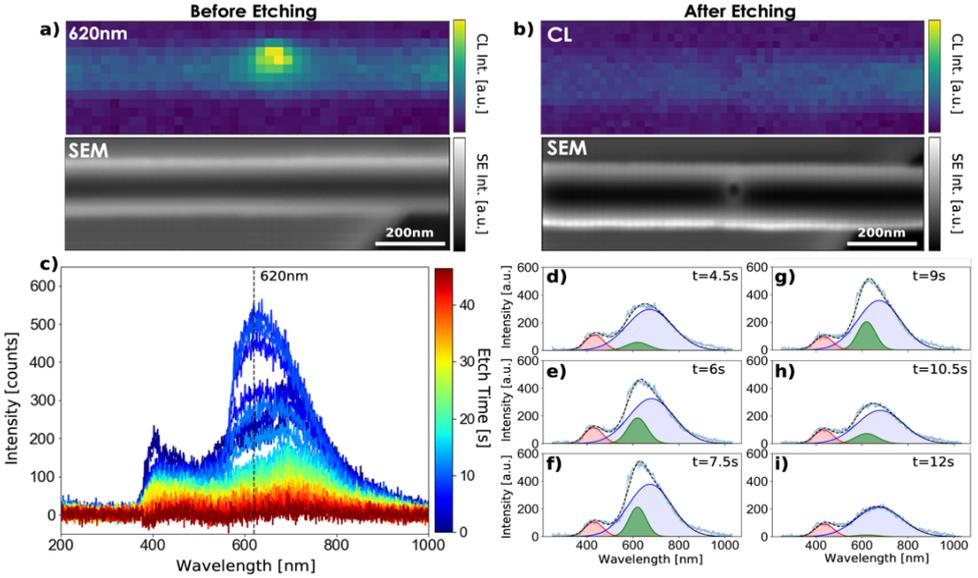

**Figure 2:** Cavity fabrication with in-situ cathodoluminescence. Hyperspectral cathodoluminescence mapping (top panel) with concurrent scanning electron microscopy (bottom panel) before etching (a) and after etching (b). (c) Time resolved cathodoluminescence with 1.5s time interval. Start and stop time is indicated by blue and red, respectively. (d-i) Individual spectra acquired from 4.5s to 12s highlighting the enhancement at 620nm. The spectra were fitted and deconvoluted with three peaks corresponding to extended defects in diamond, SnV-, and potential transition-radiation induced by the electron-beam.

### 2.2 Diamond QMC fabrication

To prepare the diamond substrate for color center creation, we etched the top 10 μm of the [001] electronic-grade single crystal diamond using Ar/Cl$_2$ plasma etching followed by O$_2$ etching to alleviate surface strain. Implantation of $^{120}$Sn ions was done at Innovion with a dose of 5 x 10$^{11}$ ions/cm$^2$ at 350 keV, targeting a depth of 76 nm as deduced from stopping and range of ions in matter (SRIM) simulations. The diamond underwent annealing at 1200°C in a vacuum furnace at 10$^{-7}$ mbar. We cleaned the diamond with a boiling mix of sulfuric, nitric, and perchloric acid to remove any graphitic residue from the annealing process. The purpose of high-dose implantation was to ensure, on average, more than one color center emitter per quantum channel.

Following the color center implantation in diamond, we coated the diamond with a silicon nitride (Si$_3$N$_4$) hard mask via plasma-enhanced chemical vapor deposition, patterned by ZEP-520A electron beam resist combined with ESpacer conductive polymer and CF$_4$ reactive-ion etching (RIE). This was followed by a conformal deposition of an 18 nm alumina atomic layer. Alumina breakthrough was achieved using CF$_4$ RIE, and an oxygen plasma was employed to undercut the diamond QMC uniformly. Finally, the silicon nitride and alumina masks were removed with hydrofluoric acid.

### 2.3 Optical measurements

We pre-characterized the diamond QMC quantum emitters using cryogenic free-space confocal photoluminescence (PL) microscopy at 4 K equipped with an objective lens (ZEISS, NA = 0.9, 100x). The collected signal is sent to either an avalanche photodiode (APD) for initial intensity mapping (Figure 1b) and for high resolution photoluminescence excitation (PLE) spectroscopy measurements, or to the spectrometer (SP-2500i, Princeton Instruments) for initial spectral analysis. Figure 1d shows a representative PLE spectrum of the SnV- quantum emitters from the chiplets. The corresponding $g^{(2)}$ characterization shows an average $(g)^{(2)}(0) < 0.1$, confirming the presence of single-photon emission. A representative autocorrelation measurement of the SnV- emitters is presented in Figure 1d. The extended optical system design is illustrated by Linsen Li et al. [28].

## 3 Results

### 3.1 Simultaneous EBIE and CL around SnV- in QMC

Figure 1c presents an SEM micrograph alongside a CL hyperspectral map of a diamond QMC obtained at an accelerating voltage of 10 keV. A spatially isolated color center was selected for the analysis to enable the independent examination of individual photon-cavity coupling and the emission characteristics of the SnV- color center. The CL hyperspectral map revealed a distinct emission peak at 619 nm, indicative of localized emission at room temperature (293 K) from the SnV- color centers. This emission manifests as a broad spectral feature, composed of several overlapping peaks. Upon cooling the system to 8 K, a pronounced enhancement of the SnV- emission peak was observed, along with a significant narrowing of the spectral full-width at half-maximum (FWHM), suggesting enhanced emission purity at reduced temperatures. Because the high-energy electron-beam excitation can excite multiple excited states that are inaccessible in resonant or near-resonant PL microscopy, no antibunching was observed in CL photon correlation function measurements, and broadband background fluorescence contaminated many of the measurements[29], [30]. However, the close correlation between the PL map shown in Figure 1b and CL maps acquired on each QMC together with the low temperature CL spectrum shown in Figure 1e provided strong confidence that the localized CL observed in Figure 1c and 1e was from a single SnV- center.

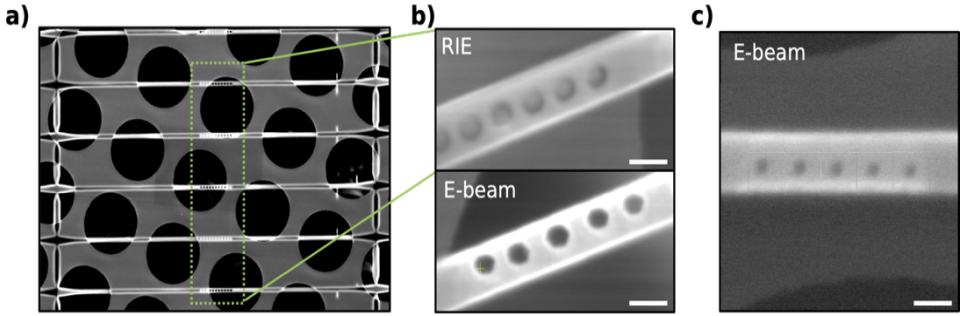

**Figure 3:** Patterning of nanophotonic structures in diamond waveguides. (a) diamond QMC suspended on holey carbon support films for E-beam in-situ fabrication. (b) photonic cavities fabricated by reactive ion etching (RIE) (top panel) and electron beam etching (bottom panel). (c) Ultimate resolution limit for periodic photonic cavities achieved with electron beam etching. Scale bars are 200 nm.

To investigate the coupling between fabricated photonic cavities and SnV- color centers, time-dependent CL measurements were performed in an a H$_2$O environment (0.75 Torr) with a beam energy of 30 keV and a current of 2.8 nA for an electron beam incident on the SnV- color center shown in Figure 2a. The resulting CL time series spectra in Figure 2c illustrate the change in SnV- luminescence as a function of electron beam etching time; the time-dependent CL and secondary electron signals are consistent with etching rates of roughly 4nm/s for these beam conditions. Interestingly, as we etch through the diamond chiplet at the location of the SnV- color center, we observed a gradual increase in the 620 nm emission. Figure 2d-I shows the spectral fitting results within the 4.5 s to 12 s timeframes to investigate the CL enhancement further. Three emission peaks were fitted and deconvoluted, where red, green, and blue correspond to diamond emission from lattice disorder (e.g. A-band emission from dislocations and grain boundaries)[31], [32], SnV- color center, and potentially other defects complexes with different charged states, respectively. The key effect observed in Figure 2d-i is the marked emission enhancement from the SnV- color center at 620 nm (green curve). This is in contrast to the red peak from the host matrix that showed almost constant emission throughout the process. Thus, EBIE can be used to dynamically control and locally optimize the coupling from a single diamond color center in a diamond photonic cavity.

While the ability to locally optimize the emission from a single color center in a prepatterned QMC using operando EBIE and CL microscopy, as shown in Figure 2 is certainly compelling, the approach described here also allows for the direct writing of nanophotonic structures in diamond QMCs around single color centers. As an initial proof of principle of this approach, the morphology of photonic cavities patterned by RIE and EBIE on this diamond QMC are compared in Figure 3. A large area image of the QMC is shown in Figure 3a, and images of the same cavity design fabricated by RIE and EBIE are shown in Figure 3b. Note, the cavities fabricated via EBIE were created by rastering the electron beam in a circular pattern to intentionally increase the cavity diameter, matching that of the RIE process. Figure 3c illustrates a similar cavity design with reduced hole diameter of 56 nm achieved by dwelling the electron beam at a single point for a user-defined time. The ability to fabricate holes and vertical cavities on these length scales provides a critical testbed for complex photon cavity designs in diamond with in-situ sub-diffraction-limited optical feedback that is not accessible with modern nanofabrication tools.

### 3.2 Origin of SnV- CL enhancement

To identify the origin of CL enhancement at 620 nm during the etching shown in Figure 2 and to better understand how the EBIE modifies the photonic local density of states (LDOS) around the emitter, we performed finite-difference time-domain (FDTD) simulations of the diamond QMC. The FDTD simulations were designed to replicate the experimental conditions as closely as possible. We modeled a SnV- embedded in a diamond waveguide, incorporating the geometric dimensions as determined by SEM and optical measurement: a width (w) of 280 nm, a height (h) of 200 nm, and a depth for the SnV- centers of 100 nm (Figure 4a). The emitter was positioned near the simulated etched dimple, offset from the center by $\delta x$ and $\delta y$ in the $x$ and $y$ directions, respectively, to account for potential misalignments (Figure 4d-f). To examine the spectral characteristics, we monitored the normalized intensity of the emission peak at 620 nm.

The FDTD simulations accounted for the emitter's assumed circular polarization and the collection optics' high numerical aperture (NA), set to 0.97, to ensure accurate collection efficiencies. By varying the etch depth, $\delta x$, and $\delta y$, we evaluated the impact of these parameters on the collection efficiency. The results indicated that the collection efficiency of the emission does not significantly change as the LDOS is modified by the etching process—less than 20% change in collection efficiency is expected for a wide range of structural parameters.

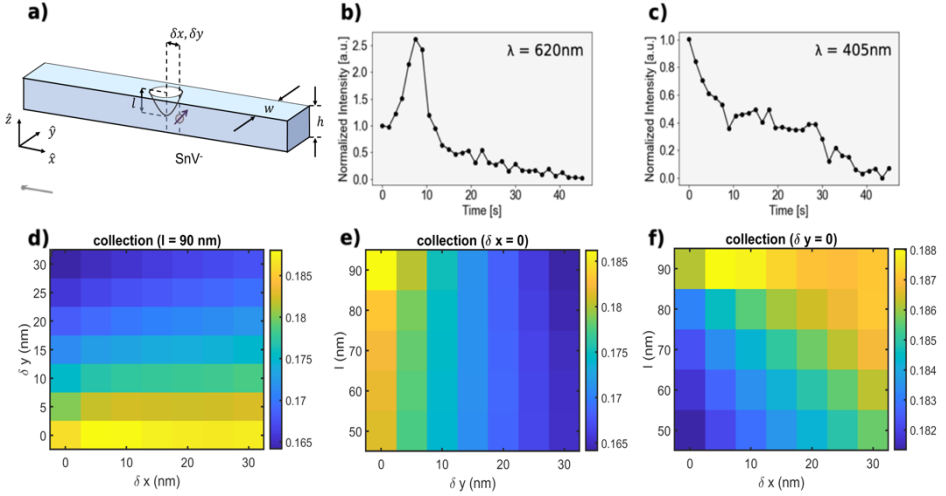

**Figure 4:** Single photon emission enhancement. (a) Illustration of an etched dimple in a diamond waveguide. Characterization with Scanning Electron Microscopy (SEM) and optical measurement revealed a width (w) of 280 nm, a height (h) of 200 nm, and a depth for the SnV- centers of approximately 100 nm. The location of the emitter is offset from the center of the etch by $\delta x$ and $\delta y$ in the $x$ and $y$ directions, respectively. (b) Normalized intensity versus electron beam etching time for emission peak at 620 nm showing over 2.5x increase in emission intensity and (c) reference peak at 405 nm. (d-f) Finite-difference time-domain (FDTD) simulation results indicate that the collection efficiency of the emitter, which depends on the etch depth, $\delta x$, and $\delta y$, is not significantly affected by the local electromagnetic environment. The emission is assumed to be circularly polarized, with the numerical aperture (NA) of the optics being 0.97.

Therefore, we conjecture that the CL enhancement is from changes in electronic excitation efficiency as a function of etching depth rather than photonic modification. Further studies are needed to identify the exact dynamics. For example, the implantation energy of Sn defects can be swept to vary the depth of the emitters and to identify the correlation with the emission intensity. Alternatively, photonic crystal cavity structures that can significantly modify LDOS can be used to see the photonic enhancement by increased LDOS[33].

## 4 Discussion

Through the precise control offered by gas-assisted EBIE, we successfully demonstrated deterministic patterning of Sn-implanted diamond QMCs, substantially enhancing the CL emission intensity. The observed enhancement in SnV- color center emission is attributed to the combination of proximity to the etched surface, LDOS enhancement, and other potential electron-emitter interactions. Further studies are needed to reveal the role of the electron-emitter interaction on the CL signal and fabrication-induced strain in diamond photonic structures fabricated using this method[36]. Tailoring the gas precursors can enable selectivity over the crystal planes and, therefore, the etching pathway (i.e., isotropic vs. anisotropic) and change the surface topology[21].

Exciting future directions include fabricating one-dimensional (1D) photonic crystals, and milling a vertical half-bowtie with a parabolic profile on (111) diamond, in which we could align the photonic mode with the emitter's polarization (polarization of SnV- is on the plane orthogonal to the defect orientation, [111])[34]. Using hermetic fiber feedthroughs to collect luminescence from fiber-pigtailed diamond QMCs could offer more refined control over light collection and delivery, ensuring that a greater proportion of emitted photons are captured and analyzed[35] while maintaining all of the advantages of in situ EBIE and CL described here.

**Acknowledgments:** The diamond QMCs fabrication was carried out in part through the use of MIT.nano's facilities. J.A. acknowledges the support from KACST-MIT Ibn Khaldun Fellowship for Saudi Arabian Women at MIT, and from Ibn Rushd Postdoctoral award from King Abdullah University of Science and Technology (KAUST). D.E. acknowledge HSBC, Mekena Metcalf, Zapata Computing, and Yudong Cao for MIT QSEC collaboration. H.C. and D.E. also acknowledge Honda Research Institute and Avetik Harutyunyan. H.C. acknowledges the Claude E. Shannon Fellowship and the Samsung Scholarship

**Research funding:** The work was supported by the U.S. Department of Energy, Office of Science, Materials Sciences and Engineering Division. The scanning electron microscopy research was performed and partially supported at Oak Ridge National

Laboratory's Center for Nanophase Materials Sciences (CNMS), a U.S. Department of Energy, Office of Science User Facility. This work was partially supported by AFOSR grant FA9550-20-1-0105 supervised by Gernot Pomrenke, the NSF Center for Quantum Networks (CQN, 1941583), and Cisco Research. J.A. acknowledges the fund from the Army Research Office MURI (Ab-Initio Solid-State Quantum Materials) Grant no. W911NF-18-1-043. H.C.

**Author contribution:** All authors have accepted responsibility for the entire content of this manuscript and approved its submission.
**Conflict of interest:** Authors state no conflict of interest.
**Data availability statement**: The datasets generated and analyzed during the current study are available from the corresponding authors upon reasonable request.